# Manipulation of Conductive Domain Walls in Confined Ferroelectric Nano-islands

*Guo Tian, Wenda Yang, Xiao Song, Dongfeng Zheng, Luyong Zhang, Chao Chen, Peilian Li, Hua Fan, Junxiang Yao, Deyang Chen, Zhen Fan, Zhipeng Hou, Zhang Zhang, Sujuan Wu, Min Zeng, Xingsen Gao,\* and Jun-Ming Liu*

Mr. Guo Tian, Mr. Wenda Yang, Mr. Xiao Song, Mr. Dongfeng Zheng, Mr. Luyong Zhang, Mr. Chao Chen, Mr. Peilian Li, Mr. Hua Fan, Mr. Junxiang Yao, Dr. Deyang Chen, Dr. Zhen Fan, Dr. Zhipeng Hou, Prof. Zhang Zhang, Dr. Sujuan Wu, Prof. Min Zeng, Prof. Xingsen Gao, and Prof. Jun-Ming Liu
Institute for Advanced Materials (IAM) and Guangdong Provincial Key Laboratory of Quantum Engineering and Quantum Materials, South China Normal University, Guangzhou 510006, China
E-mail: xingsengao@scnu.edu.cn (X. S. Gao)

Prof. Jun-Ming Liu
Laboratory of Solid State Microstructures and Innovation Center of Advanced Microstructures, Nanjing University, 210093, China

Conductive ferroelectric domain walls —— ultra-narrow and configurable conduction paths, have been considered as essential building blocks for future programmable domain wall electronics. For applications in high density devices, it is imperative to explore the conductive domain walls in small confined systems while earlier investigations have hitherto focused on thin films or bulk single crystals, noting that the size-confined effects will certainly modulate seriously the domain structure and wall transport. Here, we demonstrate an observation and manipulation of conductive domain walls confined within small $BiFeO_3$ nano-islands aligned in high density arrays. Using conductive atomic force microscopy (CAFM), we are able to distinctly visualize various types of conductive domain walls, including the head-to-head charged walls (CDWs), zigzag walls (zigzag-DWs), and typical 71° head-to-tail neutral walls (NDWs). The CDWs exhibit remarkably enhanced metallic conductivity with current of ~ nA order in magnitude and $10^4$ times larger than that inside domains (0.01 ~ 0.1 pA), while the semiconducting NDWs allow also much smaller current ~ 10 pA than the CDWs. The substantially difference in conductivity for dissimilar walls enables additional manipulations of various wall conduction states for individual addressable nano-islands via electrically tuning of their domain structures. A controllable writing of four distinctive states by applying various scanning bias voltages is achieved, offering opportunities for developing multilevel high density memories.

**1. Introduction**

Ferroelectric domain walls (FEDW),[1-2] which are often viewed as two-dimensional (2D) homo-interfaces, may possess distinctly electronic, magnetic, and optoelectronic properties localized at a 1 ~ 10 nm length scale.[3-8] The domain walls can also be created, reshaped, and displaced by external electric field,[9,10] promising for future electronic, spintronic, and optoelectronic devices. In particular, the discovery of enhanced electrical conductivity of FEDW in otherwise insulating ferroelectrics,[3] has open a new avenue for developing wall nanoelectronics, and inspired intensive research interests.

In the past decade, enhanced wall conduction has been observed in various proper and improper ferroelectrics,[3,11-15] and numerous intriguing conduction behaviors and the underlying mechanisms have been revealed.[16-27] For instance, a unique feature of quasi two-dimensional (2D) electron gas was observed in strongly charged domain walls (CDWs), which exhibit a giant metal-like conductivity up to $10^9$ times as good as the insulating bulk.[20,24] It was also reported that CDWs can be form in ultrathin ferroelectric barriers of tunnel junctions, and this approach develops discrete quantum-well energy levels leading to strong quantum oscillations in tunneling conductance.[25,26] These exciting breakthroughs have not only deepened our understanding of new physics related to FEDWs, but also pushed forward the harness of them (*e.g.* CDWs) toward practical applications in high density electronic devices. Recently, Nagarajan and Seidel *et al* proposed a two-terminal prototype of scalable non-volatile wall resistive memory with high OFF/ON resistance ratio.[28] Jiang *et al* also demonstrated a three-terminal memory device that is able to produce large reading currents along with long retention time, by utilizing retractable partially switched domain walls.[29]

In spite of these fascinating properties in association with wall conduction, major obstacles that hinder their further applications remain and reports on highly promising prototype devices are yet rare. The previously proposed device prototypes based on FEDWs in films and single crystals usually have the lateral architectures with low integration density,[28,29] and it is known that for high density scaling up technique, perpendicular architectures (*e.g.* ferroelectric/resistive random-access memory) based on nanostructures are much preferred. Besides, it is still a tricky issue to deterministically tailor the wall conduction states for on-demand applications in devices. These issues could be well addressed if the conductive FEDWs can be constrained in small nano-dots/nano-islands. Such a strategy does enable the device scaling-up to ultrahigh density, but the size-confinement and surface effects of nanostructures certainly make the wall conductiion states very different. If these additives are well understood and controlled, the FEDW conduction as a novel functionality would be offered an additional degree of freedom. In fact, recent studies did reveal some unique exotic domains such as flux-closure vortex and center-type topological states in nanoscales ferroelectrics,[30-37] which also add more ingredients into the enthralling DW functionalities. In view of the tantalizing properties in both domain wall conductance and nano-ferroelectrics, it is ofhighly promising and extremely important to look into the FEDW conduction behaviors in size-confined nanostructures. This is the major motivation of the present work.

In this work, we report the observation and manipulation of conductive domain walls confined in epitaxial $BiFeO_3$ (BFO) nano-islands (see Figure 1). The main results can be highlighted from several aspects. First, we have identified various types of conductive domain walls including zigzag-like wall, head-to-head CDW, and neutral domain wall (NDW) inside individual nano-islands. These walls exhibit markedly different conductive behaviors, like quasi-2D metallic behavior for head-to-head CDW, and semiconducting behavior for NDW.

Moreover, both the wall structures and their conduction states can be well-tailored using various external electric bias, allowing predesign control of conduction states in individual nano-islands, highly advantageous for multi-level wall conductivity memory with perpendicular device architecture.

## 2. Results and Discussion

**Fabrication of BFO nano-islands.** The BFO nano-island arrays under investigation were patterned from high quality epitaxial BFO films (~ 35 nm in thickness) via a nanosphere lithography technique using polystyrene (PS) nano-sphere template.[35] A schematic fabrication procedure is illustrated in Figure S1 (the Supporting Information). First, the well-packed monolayer of PS spheres was transferred onto the BFO epitaxial film surface, and then these nano-spheres were subjected to size shrinkages by oxygen plasma to develop discrete ordered array as template mask for patterning. Subsequently, the as-prepared product was etched by Ar ion beam, and finally the PS mask was lift-off by chloroformic solution, generating the well-ordered nano-island array, as shown in the scanning electron microscopy (SEM) image of Figure 1(a). The X-ray diffraction (XRD) θ-2θ spectrum of the sample, given in Figure 1(b), shows the good epitaxy structure, reflected by the diffraction peaks from the substrate STO (002), bottom electrode SRO (002), and BFO (002), and further confirmed by the (103) reciprocal space map (RSM) shown in Figure 1(c). We can also obtain that the in-plane and out-of-plane lattice parameters of the BFO nano-islands are $a = b = $ ~ 0.391 nm and $c = $ ~ 0.406 nm, comparable with those from in-plane compressively strained BFO.[29]

**Domain structures and wall conduction.** To see the domain wall conduction behaviors, these nano-islands were examined using conductive atomic force microscopy (CAFM) and piezoresponse force microscopic mapping (PFM), as detailed in Materials and Methods section and schematically shown in Figure 1(d). An example of the electrical conduction in these nano-islands is given by the CAFM mapping (Figure 1(e)), superimposed with its 3D AFM topology for better illustration. Clearly, some distinct zigzag-like conduction paths insides the nano-islands can be identified which are related to the conductive walls. By comparing the CAFM image with its corresponding in-plane PFM image (abbreviated as L-Pha), one could see that the most domain walls show high conductivity compared to that of domain interior (see Figure S2(c,d), supporting Information).

To careful examine the domain structures of the nano-islands, we perform a vector PFM measurement by involving both vertical-PFM and lateral-PFM scanning before and after a 90° rotation of the sample. These piezoresponse images allow us to construct the 3D piezoresponse vector contour, following the vector PFM analysis method proposed by Rodriguez and Kalinin *et al.*[38] An example for the nano-islands at pristine state is shown in Figure S4 in Supporting Information, while two representative domain states for individual nano-islands are given. The most popular state is presented in Figure S3(g) with a clear zigzag domain wall from the lateral component of polarization. The second most popular state exhibits two sets of zigzag walls with intersection to some extent (see Figure S3(h)), wherein the overlapped region forms an anti-vortex structure. The domain structures of nano-islands can be well switched by using a scanning bias of -3.5 V, as reflected by the apparent contrast change in both vertical and lateral PFM images (see Figure S4).

It is also noted that if the walls at the pristine state (with downward polarization, shown in Figure S3) exhibit faint CAFM contrast with small conduction close to the noise level (see Figure S2(a,b)). However, once a region of nano-island array was electrically poled with a -3.5 V voltage (with upward polarization), the conducting paths show much better contrast and thus much higher current density. The different conductive levels between the upward and downward polarization states of the nano-islands, is likely due to the polarization modulated resistive switching behavior. As shown in Figure S5, the domain walls with downward polarization states shows very low conduction level, in contrast to the high conductive FEDW for upward polarization states. Therefore, we will mainly focus on the domain walls in upward polarization states in the following sections.

To illustrate the one-to-one correspondence between the walls and conduction paths, both the PFM and CAFM images on a nano-island region which was previously downward-switched using a bias of -3.5 V are shown in Figure 2. Figure 2(a-c) demonstrate the PFM images scanned at angles of 0 and 90° (only L-PFM images are shown for simplicity) respectively, along with the corresponding CAFM images. Both the PFM and CAFM images are superimposed with the 3D topographic images for better recognition. From the lateral PFM images, it is revealed that most of the nano-islands show the typical zigzag domain walls, although some others do show wavy or straight head-to-head CDWs and minor amount of straight neutral domain walls (NDWs). The one-to-one correspondence between the walls and current paths (see Figure 2(a-c)) can be established, but different types of walls do have very different magnitudes in current density.

To carefully examine the wall conduction, we pick out two nano-islands with typical domain structures (marked with dashed circles in Figure 2(a-c)) for a detailed discussion. Their CAFM images together with a complete set of PFM images are illustrated Figure 2(d) and (e) respectively. The first nano-island (Figure 2(d)) has one head-to-head CDW along with one 71° NDW. The second nano-island (see Figure 2(e)) includes one typical zigzag wall, consisting of two 71° NDWs at the side edges and one very short head-to-head CDW at the zigzag corner. The different types of walls are also reflected by the dissimilar levels of conduction current density shown in the CAFM image. The current profiles across both the NDWs and CDWs are plotted in Figure 2(f), and one can distinctly identify that the head-to-head CDWs including the straight wall and zigzag wall corner show high current levels on the ~ nA order of magnitude, in contrast to the ~ pA order of magnitude for the 71° NDW. This difference is further verified by the current-voltage (*I-V*) curves measured at different locations (Figure 2(e)), which clearly indicates the largest current (> 1 nA) at the pure head-to-head CDW, intermediate current (~ 10 pA) at the NDW, and the lowest current (< 0.2 pA) inside the domains.

**Temperature dependent of conduction behaviors.** Subsequently, we discuss the conduction mechanisms of these walls by checking the temperature ($T$) dependent conduction, as shown in Figure 3. The CAFM maps for the CDW and NDW measured at various $T$, as presented in Figure 3(a-b), suggest a gradual decrease in current density for the CDW while a monotonous increase for the 71° NDW with increasing of temperature. This also implies that the CDW is electrically metallic and the NDW is semiconducting, as further confirmed by the current profiles shown in Figure 3(c-d) and the $I$-$T$ curves shown in Figure 3(e) for both types of walls. While the data for the NDW may not be sufficient for a reliable conclusion of the transport mechanism, a thermionically activated semiconducting behavior can be argued. It should be mentioned that the results reported here are of generality and applicable to almost all nano-islands of our samples. The metallic conduction of the head-to-head CDW is somewhat similar to earlier observations on $BaTiO_3$ single crystals and La-doped BFO films.[20,24]

Currently, there have been proposed several mechanisms to interpret the wall conductivity, including the band bending, defects, and local lattice distortion among others, noting that no general agreement so far has been reached.[16-20] Here, the identified metallic conduction of these CDWs can be associated with the quasi-2D electron gas of these walls due to the effective compensation of bound polarization charges by the accumulated charges.[20] This can also be understood by the band bending at the CDW induced by the uncompensated polarization charges, which leads to a dramatic drop of the conduction band below the Fermi-level, thus resulting in the metallic behavior. In contrast, only a small drop in conducting band occurs at the NDW, leading to no more than the slightly enhanced conductivity.

**Manipulation of wall conduction states.** Given the established one-to-one correspondence between the domain wall type and conductivity, it is proper to address how the wall conduction state can be manipulated by switching the wall from one type to another, which is key for the promising data memory devices. The capability to manipulate the walls and their conduction states by electric field is the most straightforward and simple approach. For realizing this target, various bias-voltage scanning trials were performed on these nano-islands, and the main results are highlighted in Figure 4.

The domain structure and wall conduction for a nano-island at the pristine state (no electrical bias) is shown in Figure 4(a): typical zigzag walls consisting of 71° NDW and tail-to-tail CDWs, with rather low wall conduction current (< 2 pA) over the whole nano-island. In addition, an electrical bias can change the domain structure remarkably in the repeatable way and thus the domain wall conduction state, allowing programmable and pre-designable control of conduction states in individual nano-islands. To illustrate this step-like/processable control, we present the results for several different events. First, a small bias voltage of -3.5 V switches the vertical polarization component upwards completely, while the lateral component remains the similar zigzag wall pattern, as shown in Figure 4(b). The new zigzag wall mainly consists of NDWs and head-to-head CDWs at the angle corners (similar to Figure 2(e)), in which the NDW exhibits a current level of ~10 pA and the CDW allows a current of ~100 pA, as mentioned previously. It is worth to noting again that the observed different conduction states for the two "zigzag" domain states shown in Figure 4(a) and (b) can be attributed to the strong resistive switching behavior between the upward and downward vertical polarization states in individual nano-islands, as illustrated in Figure S5.

Second, if the bias voltage is -4.5 V instead of -3.5 V, as shown in Figure 4(c), one sees that the zigzag wall now converts to a straight head-to-head CDW, which displays the largest conductive current (~ nA over the whole long wall). Third and more interestingly, if the bias voltage is further enhanced up to -5.5 V, as shown in Figure 4(d), it is surprising to see that the CDW vanishes and it converts to pure stripe-pattern consisting of typical 71° NDWs, leading to a sudden drop in the current (~10 pA).

The above example suggests that the both the domain wall pattern and conduction state for a nano-island can be deterministically controlled by external electric bias, giving rise to four different conduction states: (1) the initial zigzag-DW state with downward vertical component of polarization, exhibiting smallest current < 2 pA；(2) pure 71° NDW state with upward vertical polarization after applying -5.5V, corresponding to intermediate state with current ~10 pA；(3) zigzag-DW state with upward vertical component of polarization, created by applying scanning bias of -3.5 V, showing a larger conduction (10 pA to 100 nA)；(4) head-to-head CDW state with upward vertical component of polarization, after applying -4.5 V, showing large current > 1nA. The observation of various conduction states in individual nano-islands clearly indicate that the conductive domain walls can be engineered and utilized in multilevel FEDW memory devices, and the schematics of the conceptual devices are illustrated in Figure 4 (e) as one proposal, in which each individual nano-island is able to store two data bits by using the four programmable conduction states. Such a device has the perpendicular architecture, compatible with ultra-high density scaling up techniques, a noticeable advantage.

To find out the driving forces for electric control of different conductive domain wall states, in particular the CDW, we perform a scanning Kelvin potential microscopy (SKPFM) measurement to probe the surface potential distribution on a nano-island. Figure 5 show surface potential (SKPFM) and corresponding CAFM maps for nano-islands with various conductive domain wall states. Among them, the nano-island carrying a head-to-head CDW exhibits the lowest surface potential, and the region adjacent to the CDW also shows a relatively low surface potential in contrast to the rest region in the same nano-island (see Figure 5(b)). This suggests that the bias-induced electron injection and trapping, reflected by the lower surface potential, likely responsible for the CDW formation. The trapped electron charges can compensate the unbalanced polarization bond-charge at the CDW and lower the formation energy, and simultaneously assist with the external bias field to trigger the nucleation CDW. At the same time, the electron charges at CDW could also lead to dramatic lowering of the conducting band, thus leading to the high conductive metallic CDWs.

On the other hand, for the nano-islands with less trapped charge density, *i.e.* relatively higher surface potential, zigzag state or 71°-NDW with small portion of CDW can be favored, see Figure 5(c,d). This further support that the above highlighted domain wall states (shown in Figure 4) could also be driven by different levels of charge injection/trapping, which can be easily modulated by external bias. As the domain wall patterns can greatly determine the domain wall conduction state in a nano-island, our observation provides a simple route to tailor the conduction states of individual nano-islands, via electrical tuning of various domain wall states.

## 3. Conclusions

In summary, we have observed various types of conductive domain walls confined in high density array of BFO nano-island, including CDWs, head-to-tail NDWs, and unique zigzag-DWs, which also exhibit much different conductive properties. The head-to-head CDWs demonstrate a metallic conductive behavior and a high conduction level up to two orders in magnitude that of the NDWs and four orders that of domain interior. The dissimilar conductive behaviors also enable the manipulation of various conduction states (*e.g.* four different conductive levels) in individual nano-islands, via electric modulation of their domain wall patterns by applying various electric bias. This creates a new avenue for further engineering the conductive domain walls in ultrahigh density DW devices, *e.g.* multilevel nonvolatile memory with perpendicular architecture.

## 4. Experimental Section

*Fabrication of nanodot arrays.* The fabrication procedure for the nano-island arrays have been illustrated in the schematic flowchart in Figure S1 in Supporting Information, which is based on nanosphere patterning on well-epitaxial BFO thin films. In brief, the epitaxial

BiFeO$_3$ thin film of ∼ 35 nm in thickness, alone with a ∼ 20 nm-thick epitaxial SrRuO$_3$ layer were deposited on the (100)-oriented SrTiO$_3$ substrates by PLD using a KrF excimer laser (wavelength $\lambda$ = 248 nm) at 680 °C with an oxygen ambient of 15 Pa, a pulse energy of 300 mJ, and a repetition rate of 8 Hz. Then, the nanoscale polystyrene spheres (PS) dispersed in a mixture of ethanol and water were then transferred onto the as grown BiFeO$_3$ film to form a close-packed monolayer template. The nanospheres are then etched to the desired size by oxygen plasma to form a discrete ordered array. This is followed by Ar ion beam etching with appropriate etching time. Finally, the PS layer is removed by chloroformic solution and the periodically ordered BFO nano-island arrays are obtained. After the patterning, the samples were also annealed at oxygen ambiance at 400 °C for 30 mins to reduce the defect and residual strain.

*Ion beam etching process.* The samples are etched by Ar ion beam etching at a vacuum pressure of 8.0 × 10$^{-4}$ Pa at room temperature. During the etching, the incident ion beam is perpendicular to the sample surface. The etching parameters have been carefully optimized, using a cathode current of 15.7 A, an anode voltage of 50 V, a plate voltage of 300 V, an ion accelerating voltage of 250 V, a neutralization current of 13 A, and a bias current of 1.2 A.

*Microstructural characterizations.* The structure of nanodots is characterized by XRD (PANalytical X′Pert PRO), including θ-2θ scanning and RSM along the (103) diffraction spot. The top view surface images are obtained by SEM (Zeise Ultra 55), and the topography images are taken by AFM (Asylum Cypher AFM).

*PFM and CAFM characterizations.* The ferroelectric domain structures of these nano-islands were characterized by PFM (Cypher, Asylum Research) using conductive probes (Arrow EFM, Nanoworld). The local piezoresponse loop measurements are carried out by fixing the PFM probe on a selected nano-island and then applying a triangle square waveform accompany with ac driven voltage, via the conductive PFM probe. The vector PFM function of our AFM unit allows simultaneous mapping of the vertical (out-of-plane) and lateral (in-plane) amplitude and phase signals from the nano-island one by one. To determine the 3D domain structures, both the vertical and lateral PFM images were collected for sample rotation at 0° and 90°. Before the rotation a specific position of the sample was artificial marked, so that we are able to track the sample region after rotation. The CAFM maps and current-voltage (*I-V*) measurement are characterized by using platinum and diamond coated probes (CONTV-PT of Bruker and CDT-NCHR-10 of Nanowold, respectively).

**Acknowledgements**

The authors would like to thank the National Key Research and Development Program of China (No. 2016YFA0201002), the State Key Program for Basic Researches of China (No. 2015CB921202), the Natural Science Foundation of China (Nos. 11674108, 51431006, 51272078), the Project for Guangdong Province Universities and Colleges Pearl River Scholar Funded Scheme (2014), the Natural Science Foundation of Guangdong Province (No. 2016A030308019), and the Science and Technology Planning Project of Guangdong Province (No. 2015B090927006).

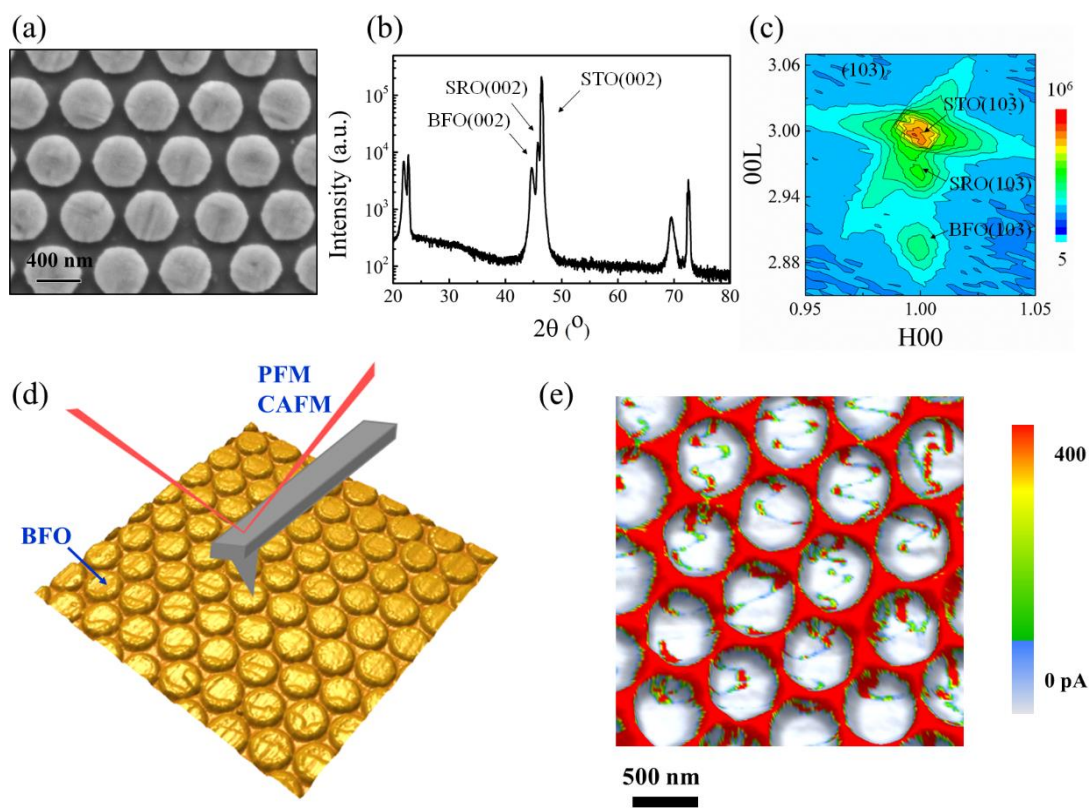

**Figure 1.** Structures and domain wall conductivity of an array of BFO nano-islands. (a) SEM image, (b) XRD diffraction pattern, and (c) reciprocal space map of nano-islands. (d)

Simplified schematic diagram of PFM and CAFM characterization of the nano-island array, and (e) an example of CAFM map of conductive domain walls in an array of nanodots, superimposed with their corresponding 3D topographic image.

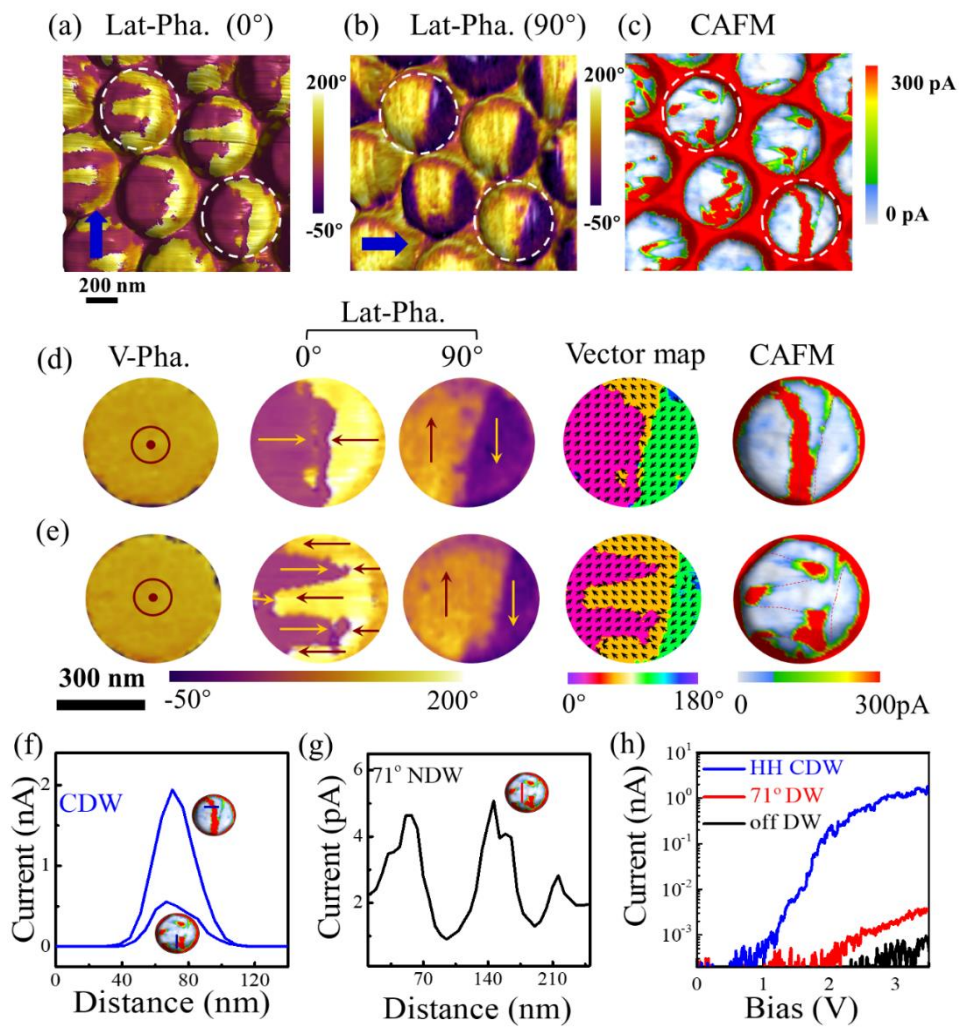

**Figure 2.** Domain structure and corresponding domain wall conductivities for an array of nano-islands. (a,b) The lateral PFM phase images captured at sample rotation angles of 0° (a) and 90° (b), respectively, which was also superimposed with their 3D topographic image for better illustration. (c) The corresponding CAFM images showing enhanced conductions at domain walls. (c,d) The domain states and their corresponding CAFM maps for two selected nano-islands with different types of domain walls: a head-to-head CDW and a NDW (d), and zigzag domain wall (e). (f,g) Corresponding current profiles extracted from (d and e) showing different domain wall current intensity for both the 71° NDW and CDW, respectively. (h) Local I-V curves on CDW, NDW, and domain interior region.

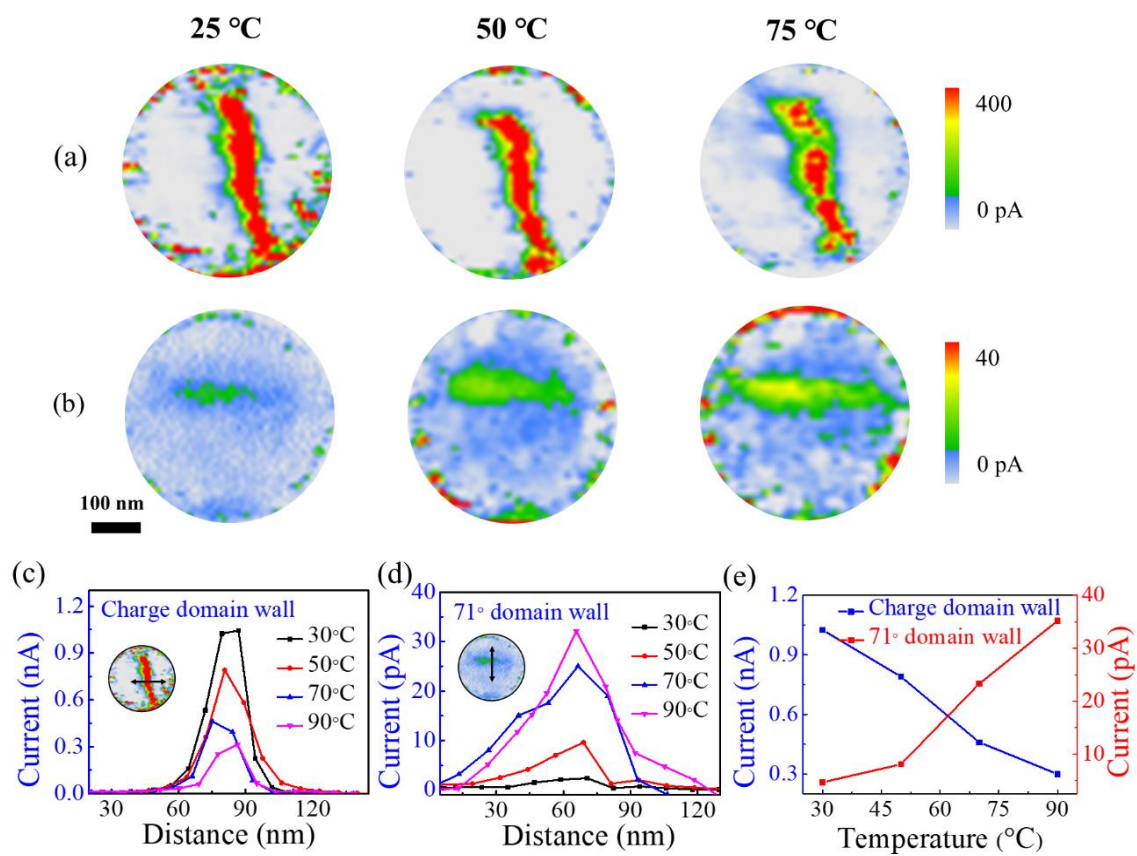

**Figure 3.** Temperature dependent of conductive behaviors for two different types of domain walls. (a-b) The CAFM maps at various temperatures for two nano-islands respectively contenting a CDW (a) and a NDW (b). (c,d) Current profiles as function of temperature for the two types of domain walls. (e) Temperature dependent conductive current curves for the two different types of domain walls

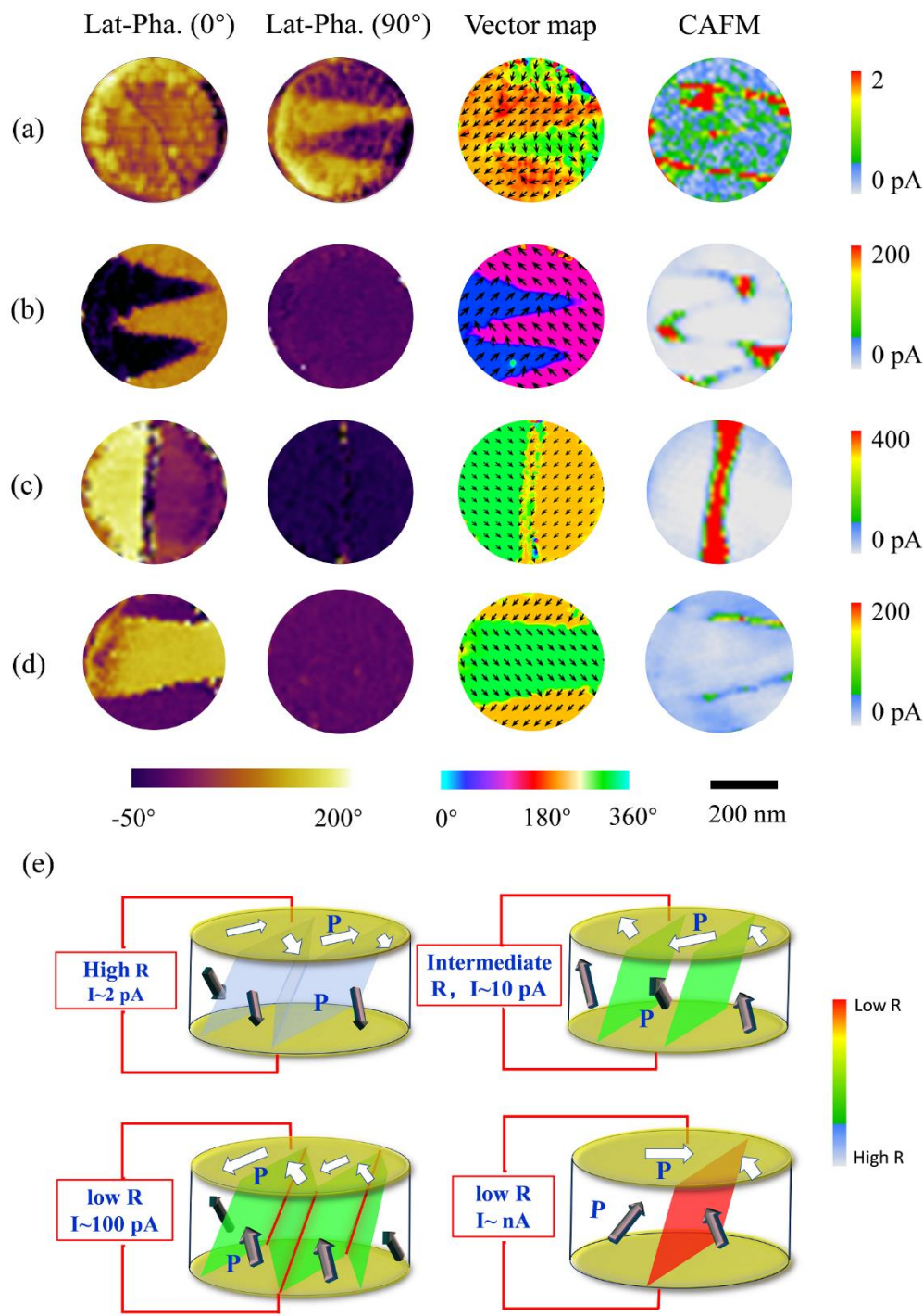

**Figure 4.** Manipulation of various domain wall patterns and corresponding conduction states in individual nano-islands by using external scanning bias voltages. (a-d) PFM and CAFM images for the four different types of domain wall states: (a) Zigzag-DW state with downward vertical polarization (pristine state); (b) Zigzag-DW state with upward vertical polarization (after poling by scanning bias voltage of -3.5 V); (c) Head-to-head CDW state with upward

vertical polarization (bias voltage -4.5 V); (d) Purely NDW state with upward vertical polarization (bias voltage -5.5 V). (e) Schematic diagrams of conceptual domain wall device utilizing the four domain wall states, which allows store two data bits in one individual nano-island cell.

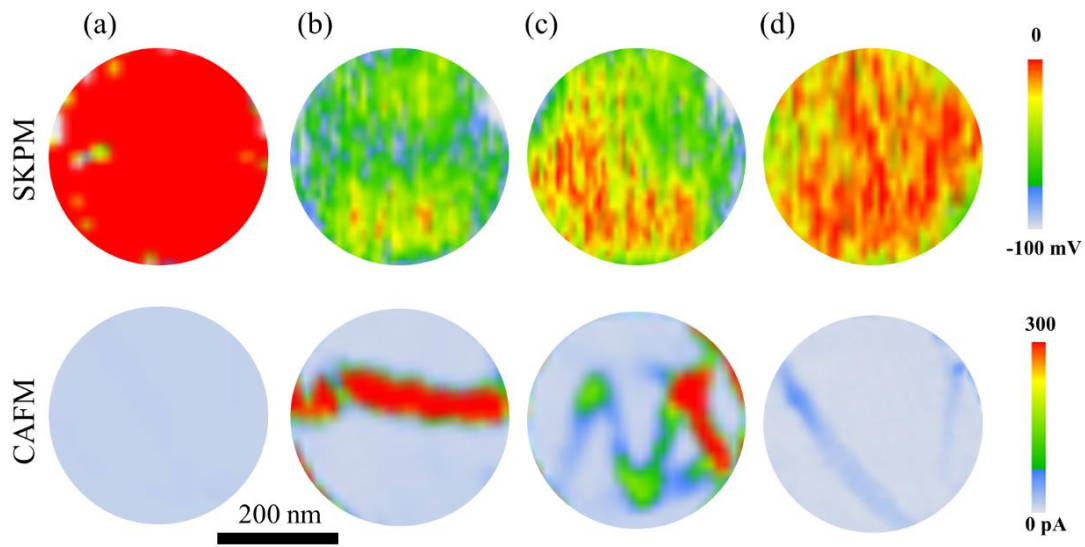

**Figure 5.** Surface potential for various domain wall state of individual nano-islands. (a-c) SKPFM (upper column) and corresponding CAFM (lower column) maps of individual nano-islands at different domain wall states: initial states with downward vertical polarization (a), charge domain wall state with upward polarization state (b), zigzag domain wall states with upward polarization state (c), and pure NDW states with upward polarization state (d).

vertical polarization (bias voltage -4.5 V); (d) Purely NDW state with upward vertical polarization (bias voltage -5.5 V). (e) Schematic diagrams of conceptual domain wall device utilizing the four domain wall states, which allows store two data bits in one individual nano-island cell.

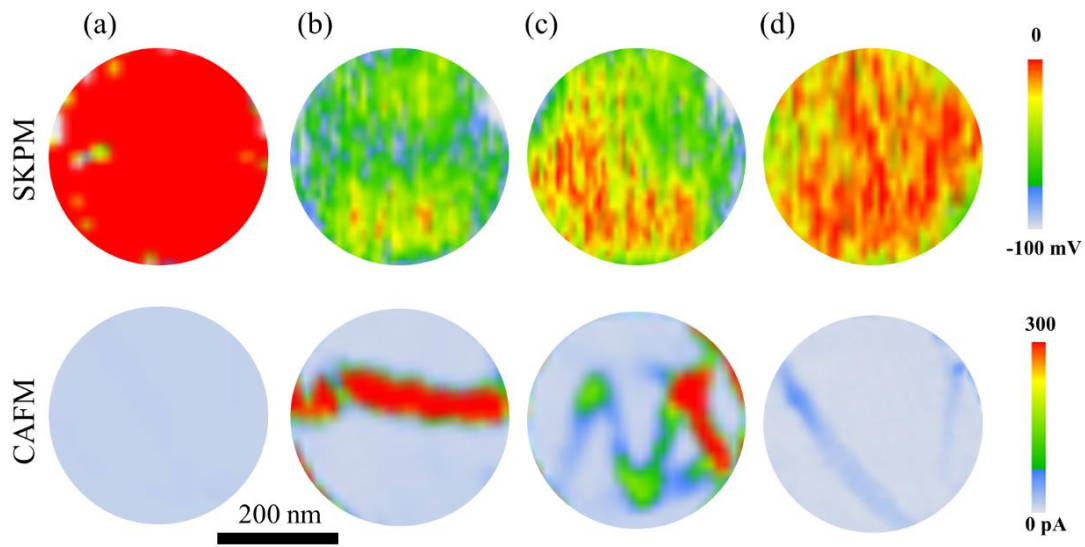

**Figure 5.** Surface potential for various domain wall state of individual nano-islands. (a-c) SKPFM (upper column) and corresponding CAFM (lower column) maps of individual nano-islands at different domain wall states: initial states with downward vertical polarization (a), charge domain wall state with upward polarization state (b), zigzag domain wall states with upward polarization state (c), and pure NDW states with upward polarization state (d).

**The table of contents entry**

High-density ferroelectric nano-island array can be repeatedly and highly reliably produced, where each island contains certain domain walls whose types and shapes can be well controlled by electric bias, and the wall conduction states can also be switched from one to another in a programmable sequence. This finding make it possible to electrically control the conductive states through domain wall engineering.

**Keyword:** Ferroelectric domain wall conductivity, domain wall memory, ferroelectric nanostructures

**Authors:** *Guo Tian, Wenda Yang, Xiao Song, Dongfeng Zheng, Luyong Zhang, Chao Chen, Peilian Li, Hua Fan, Junxiang Yao, Deyang Chen, Zhen Fan, Zhipeng Hou, Zhang Zhang, Sujuan Wu, Min Zeng, Xingsen Gao,\* and Jun-Ming Liu*

**Title:** Manipulation of Conductive Domain Walls in Confined Ferroelectric Nano-islands

**ToC figure**

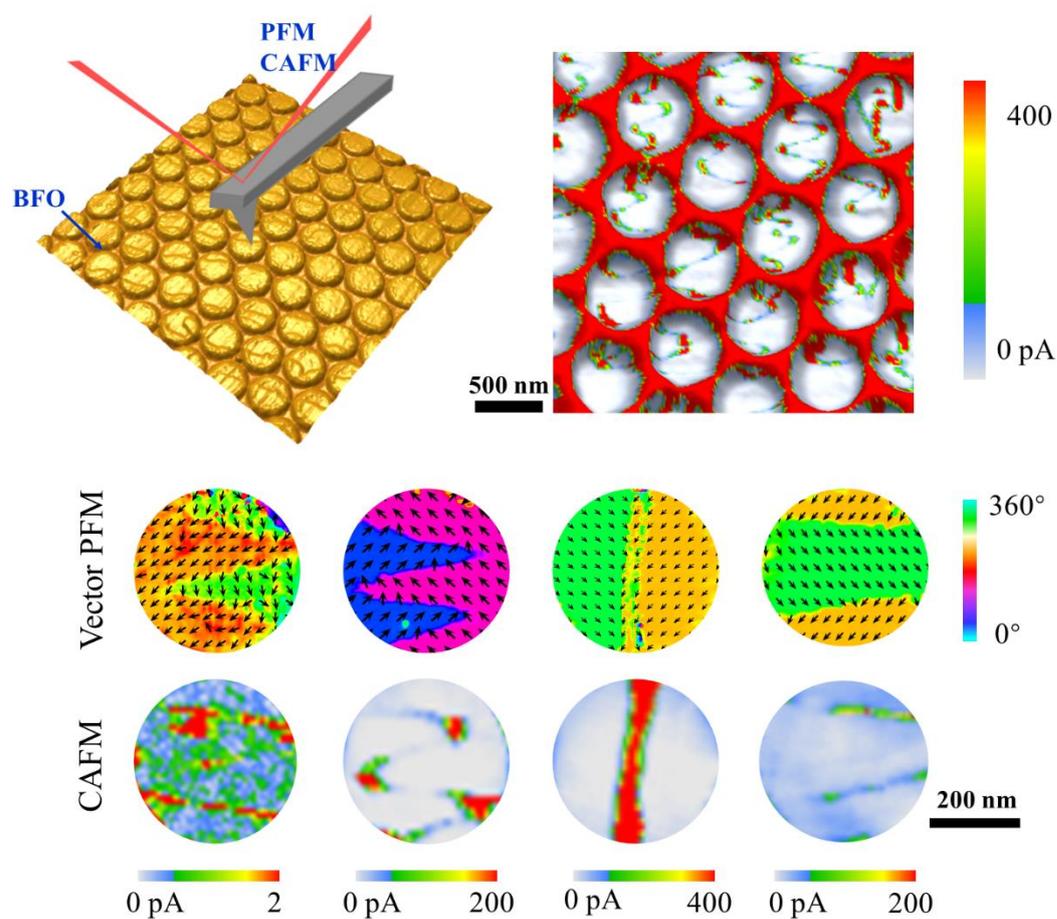

# Supporting Information

**Title:** Manipulation of Conductive Domain Walls in Confined Ferroelectric Nano-islands

**Authors:** *Guo Tian, Wenda Yang, Xiao Song, Dongfeng Zheng, Luyong Zhang, Chao Chen, Peilian Li, Hua Fan, Junxiang Yao, Deyang Chen, Zhen Fan, Zhipeng Hou, Zhang Zhang, Sujuan Wu, Min Zeng, Xingsen Gao,\* and Jun-Ming Liu*

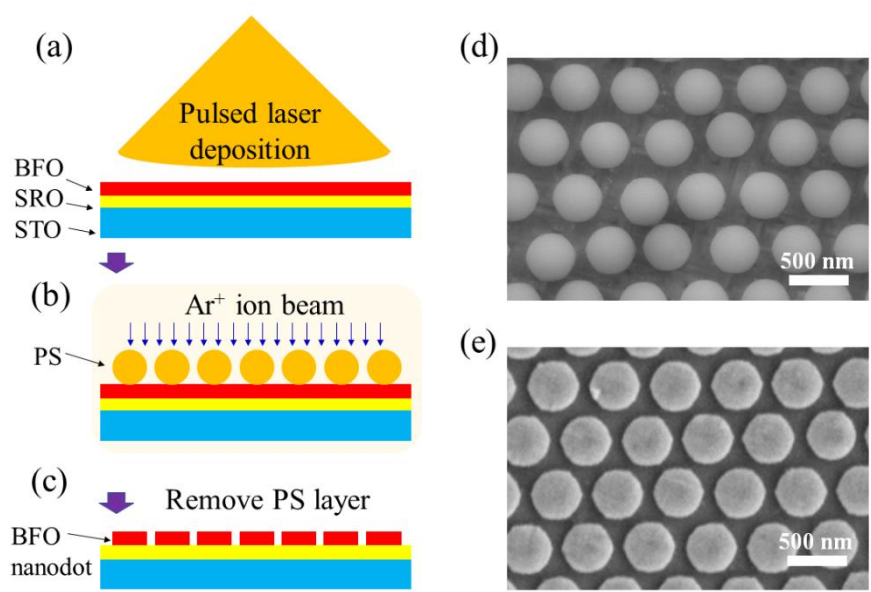

**Figure S1.** Fabrication details for nano-island arrays. (a-c) Schematic flowchart illustrating the fabrication procedures of BFO nano-island arrays on STO substrate: Pulsed laser deposition (a); Ar ion beam etching through PS nanosphere template layer (b); Removal of PS template (c). (c,d) The corresponding SEM images of discretely ordered monolayer PS nanospheres which were previously subjected to size shrinkage by oxygen plasma etching (d), and the ordered BFO nano-island arrays after Ar ion beam etching and PS lift-off (e).

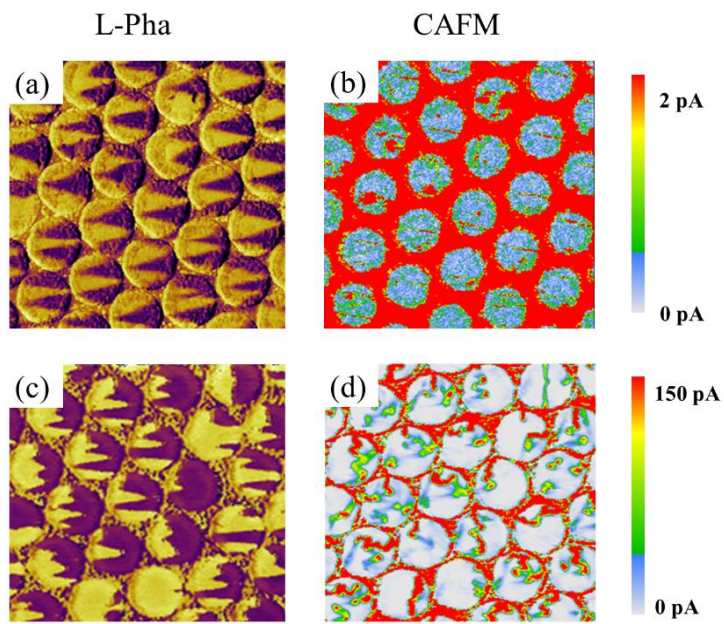

**Figure S2.** PFM and CAFM images for a nano-island array at (a,b) Lateral PFM image and CAFM image for an array of nano-islands at pristine state with vertical polarization pointing downwards. (c,d) Lateral PFM image and CAFM image of a region of nano-islands, which have been poled upwards by using a scanning bias voltage of -3.5 V.

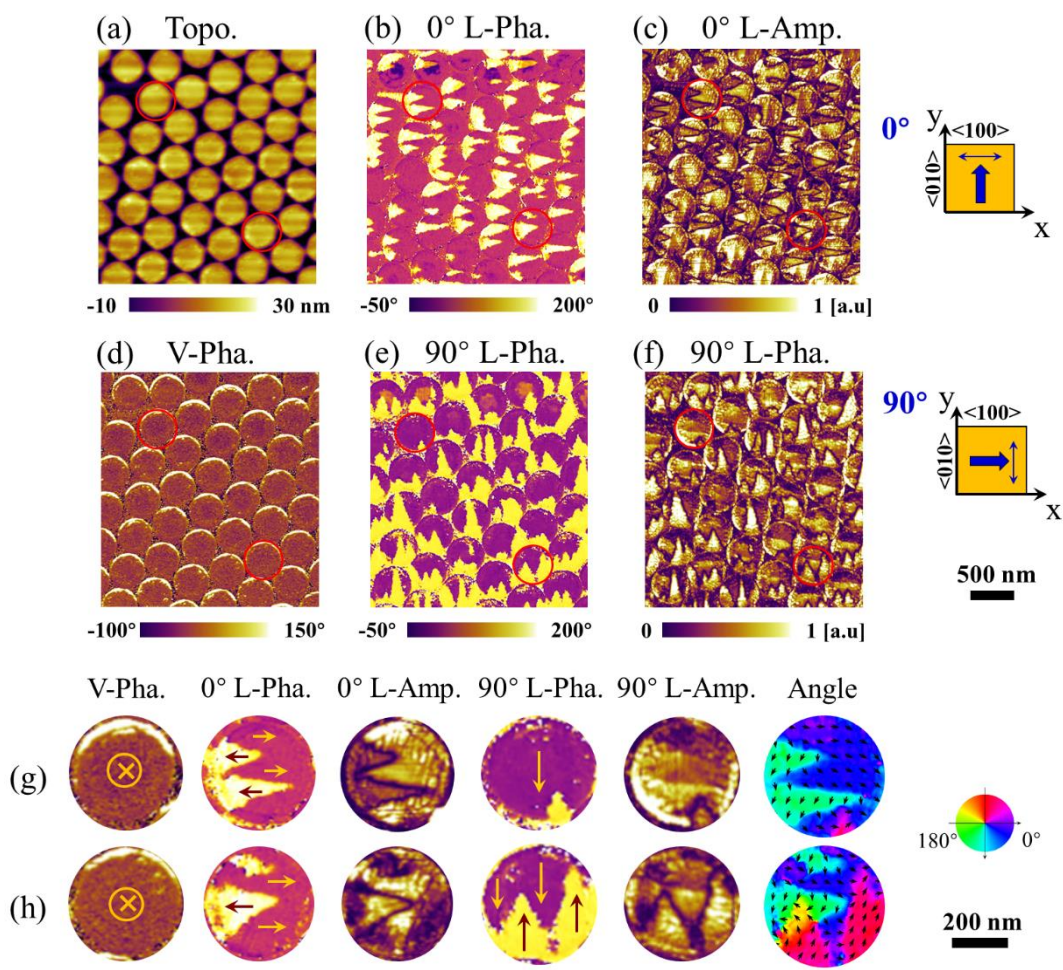

**Figure S3.** Vector PFM images of an array of nano-islands. (a-f) The topography (a), vertical PFM phase image (d), and lateral PFM phase and amplitude images captured with sample rotation for 0° (b,c), and with 90° (e,f). (g,h) The PFM vector maps of two frequently observed domain states for individual nano-islands: with a zigzag domain wall (g), with an anti-vortex domain (h).

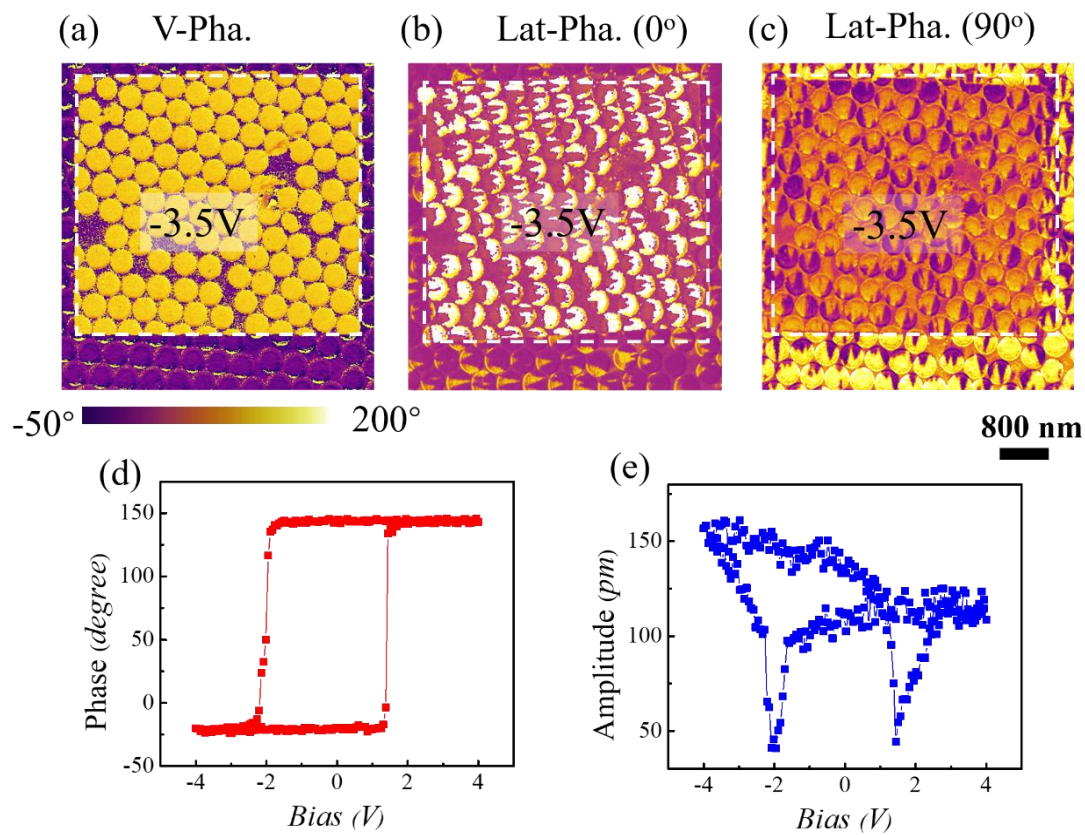

**Figure S4.** Polarization switching for an array of BFO nano-island. (a-c) Vertical PFM phase image (a), and lateral PFM phase images captured for 90° and 0° orientation (b,c), in which the middle square area was poled upwards (with a tip bias voltage of -3.5 V), and the rest nano-islands remain downward polarization. Local piezoresponse hysteresis loops acquired on a single nanodot: the phase-voltage (d) and amplitude-voltage piezoresponse (e) hysteresis loops.

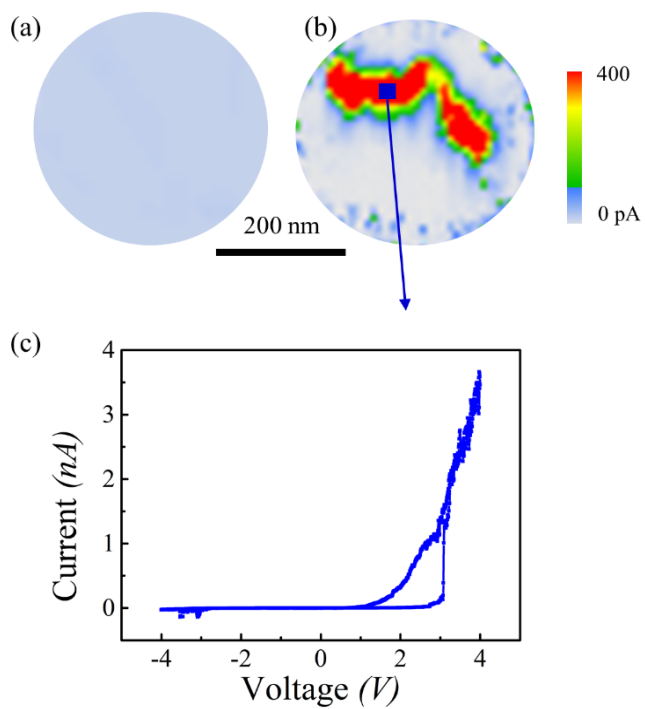

**Figure S5.** Polarization modulated resistive switching behavior of a nano-island. (a,b) Comparison of conductive CAFM maps between downward (a) and upward (b) polarization states of the nano-island. (c) I-V curve indicating apparent resistive switching between upward and downward polarization states.